\documentstyle{article}
\begin{document}                                                               
\vspace{1cm} 
\centerline{ \Large  
{\bf  
Two interacting diffusing particles on low-dimensional discrete structures}}
  
\vspace*{2cm}  
\centerline {\large   
Raffaella Burioni${}^{1}$,   
Davide Cassi${}^{1}$,  
Giovanni Giusiano${}^{2}$,   
Sofia Regina${}^{1}$  }
  
\centerline{\it Istituto Nazionale di Fisica della Materia - INFM}  
\centerline{${}^1$\it Dipartimento di Fisica,  
Universit\`a di Parma, Parco Area delle Scienze 7A, 43100 Parma, Italy }  
\centerline{${}^2$\it Dipartimento di Fisica,  
Universit\`a di Perugia, Via A. Pascoli, 06123 Perugia, Italy }  
\vskip 1truecm  
\begin{abstract}   
In this paper we study the motion of two particles diffusing on   
low-dimensional discrete structures in presence of a hard-core   
repulsive interaction. 
We show that the problem can be mapped in two decoupled problems of
single particles diffusing on different graphs by a transformation we call
{\it diffusion graph transform}. This technique is applied 
to study  two specific cases: the narrow comb and the 
ladder lattice. 
We focus on the determination of the long time probabilities for  the 
contact between particles and  their reciprocal crossing.
We also obtain the mean square dispersion of the particles 
in the case of the narrow comb lattice.
The case of a sticking potential and of   
`vicious' particles are  discussed.

\end{abstract}  
\vskip 1.5truecm  
%UPRF-97-10  
\newpage  
\section{Introduction}  
The diffusion of interacting particles has recently become the object  
of a renewed interest due to the great variety of its applications [1-14].  
The first systematic approaches to the problem were introduced in 
mathematical literature by Harris \cite{harris} and then by Fisher in  
\cite{Fish}, where the simultaneous motion of $p$ random walkers was studied 
and multiple occupation of a single site was forbidden by the hard 
core repulsion.  
Since then, different kinds of interactions have been considered,   
like the so called `\textit{n-}friendly' walkers   
(two of the  walkers can move together for up to $n$ lattice sites) and 
the  
`vicious' walkers (walkers kill each other when they meet and the diffusive   
process stops) \cite{Fish,Arrow,Wall}. \\   
In general, the case of two diffusing interacting particles can be regarded as the motion 
of one particle moving on a randomly evolving environment. This gives rise to highly 
non linear behaviors and implies a strong influence of the underlying geometry on the  
process. This has already been evidenced in the above mentioned 
works, even if only standard $d$-dimensional lattices have been considered. 
 
Interactions play a dramatic role in low dimensional systems, where 
multiple collisions between particles are so frequent that they completely 
modify the usual random walks behavior.  
Extensive studies have been carried on in the simplest case of $1$-dimensional lattices  
where anomalous behaviors have been evidenced  \cite{Asl1}. 
Now since relevant applications of diffusion with interactions concern biological 
systems, complex networks and abstract spaces which typically exhibit low dimensionality 
but non standard geometry, the usual modeling with regular lattices is 
not appropriate. These structures 
are naturally represented by graphs and their study requires new mathematical approaches  
based on graph theory. 
 
In this paper, we focus on the diffusion process of two particles in presence of a  
contact repulsive interaction on low-dimension structures of non conventional geometry 
called the {\it `Narrow Comb`} and the {\it `Ladder`} lattice.  
These graphs reproduce the geometrical feature of some simple polymers, like polymeric
liquid crystals, and decorated linear networks. 
Though keeping the same large scale structure of a linear chain, they  
allow particle crossing in spite of the presence of a repulsive interaction. 
 
The problem of two diffusing particles on these graphs is solved introducing  
a new technique we call {\it the diffusion graph transform} (DGT)\cite{tesi}. 
The original   
problem can be decoupled into the motion of two independent random walkers 
moving on  
two graphs, representing the coordinate   
of the center of mass and the relative distance between the two particles 
respectively.  
Then the problem can be analytically solved using discrete time random walks methods on  
non-translation invariant lattices \cite{MeW,Bethe}.  
Moreover, modifications of  
the potential (sticking potential or vicious walkers potential) can be easily taken  
into account.\\  
The paper is organized as follows. In section $2$ we present DGT 
in the simple case of the regular $1$-dimensional chain. 
In section $3$ we study and solve the case  
of the Narrow Comb while on section $4$ we consider the Ladder graph. 
Finally in section $5$ we discuss our results. 
 
\section{The diffusion graphs}  
  
As a first example of application of the diffusion graph technique, we present 
the simple and well studied case of the linear chain.  
Let us consider a \textit{1-d} lattice with two  
classical distinguishable particles with coordinates $x_{1}$ and $x_{2}$.  
At $t=0$ the particles start from two adjacent sites  
$x_{1}=0$ and $x_{2}=1$. At each discrete time step they can jump to  
one of their nearest-neighbors with the same probability  
$p=\frac{1}{2}$ if these sites are unoccupied.    
When the particles are on two adjacent sites, they are forced to move   
apart and to occupy their only empty nearest neighbor: these prescriptions   
model the hard-core repulsive interaction. One of the main features on the linear chain is 
that the two particles cannot cross each other and therefore $x_{1}<x_{2}$.\\
The key step in the diffusion graph technique is the introduction of  
the coordinate of the center of mass  
\begin{equation}  
	{c_{m}}  =  {\frac{x_{2}+x_{1}}{2}}-\frac{1}{2}  
	\label{uno}  
\end{equation}  
and the value of the relative distance  
\begin{equation}  
	{d_{r}}  =  {\frac{x_{2}-x_{1}}{2}}-\frac{1}{2}  
     \label{due}  
\end{equation}  
where the additional term $-\frac{1}{2}$ defines the initial conditions  
$c_{m}=d_{r}=0$. \\  
At each discrete time step, the value of $c_m$ can change to $c_m+1$ and $c_m-1$,  
corresponding to both particles moving   
to the right or to the left, or can remain unchanged if particles move in opposite directions.  
The probability of these moves can be easily calculated from the original  
process rules and they can be shown to be respectively $p=1/4$ and $p=1/2$.  
An analogous derivation can be obtained for the value of $d_r$, leading to 
the  
same conclusions, with the only additional prescription that $d_r$ cannot 
assume negative  
values. This suggest to represent the value of $c_m$ and $d_r$ as the positions of two random  
walkers moving on two new lattices  we shall call the {\it diffusion graphs}.\\  
In the case of  $c_m$ the diffusion graph will be a linear chain with jumping probabilities 
$p=1/4$ and waiting time probabilities in each site  $p=1/2$   
as represented in Fig.1 a). The diffusion graph of $d_r$,  
represented in Fig.1 b),   
is a half linear chain with  waiting time probabilities $p=1/2$  
in every site, except in the origin $O$, since this point represents the 
contact between the 
particles ($d_r=0$). Once the diffusion graphs are defined, the original problem is reduced 
to the study of a single random walker on the diffusion graphs of   
$c_m$ and  $d_r$ respectively.\\
In each single walk the time evolutions of  $c_m$ and    
$d_r$ are correlated in such a way that, when one of these coordinates 
changes, the other is left unchanged.
However this very particular kind of correlation has no influence on the average values
since it is completely taken into account by the waiting time probabilities on each
diffusion graphs. 
The only residual correlation appears in a modified waiting time probability
when the two particles have a contact, since in that case 
$d_{r}$ must increase of one unit, with a corresponding waiting time 
probability $p=1$ for $c_m$ instead of the usual $p=1/2$. 
However it can be proven that
this change in the waiting probability does not affect the leading asymptotic 
behaviors for $t\to\infty$ \cite{tesi}.\\ 
By standard random walks technique on the diffusion
graphs, one obtains
the relevant quantities for two interacting particles on a chain.  
The probability of a contact between particles is mapped into the probability of 
returning to point $O$ for a random walker on the $d_r$ diffusion graph and 
follows the asymptotic law: $P_{O}(t)\sim \frac{1}{2\sqrt{\pi}} t^{-1/2}$ as 
$t \rightarrow \infty$.  
The coordinate of each particle and of the center of mass grows as $t^{1/2}$. 
\cite{Asl1}. We can also consider a sticking probability after a contact simply   
by introducing a waiting time probability in point $O$ of the $d_r$ diffusion   
graph. Finally, the problem of `vicious particles', killing each other when   
occupying nearest neighbors sites, can be described by considering only the probabilities of  
reaching $O$ for the first time on the diffusion graphs.  
  
\section{Two interacting particles on the narrow comb lattice }  
The possibility of particles crossing is introduced on the linear chain by  
adding fingers of unit length at each site. The corresponding geometrical
structure is the  
narrow comb lattice (NCL) represented in Fig. 2 a.\\  
Let us consider two particles starting at $t=0$ from the adjacent sites $x_1=0$   
and $x_2=1$ on the backbone of the comb. The particles move on the NCL according to 
the following rules:\\  
{\it i)} each particle can jump from site $i$ to one of its nearest neighbors sites 
with probability $1/z_i$, $z_i$ being the number of nearest neighbors of site $i$. 
This means that when a particle is on one of the fingers, it is forced to 
jump to 
the backbone with probability equal to one.\\  
{\it ii)}  When the two particles occupy two adjacent sites on the 
backbone (contact),
the particles are forced to move in one of the following ways:\\
{\it a)} the two particles jump on the backbone following  opposite 
directions\\
{\it b)} both particles jump on the fingers corresponding to their positions\\
{\it c)} only one of the particles jumps on the finger while the other moves on the backbone. 
This can be done in four different ways. Two of them lead to what we shall call 
a `tower configuration', where the 
particles occupy a site on the backbone and the site of the corresponding finger.
\\ 
{\it iii)} the only allowed evolution of a tower configuration consists in the 
lower particle taking a further step on the backbone, while the particle 
on the finger drops down to the 
backbone. This means that particles have a probability 1/2 to cross each 
other, depending on  
the lower particle taking the step to the left or to the right.\\  
Now we can build the diffusion graphs for the center of mass $c_m$ and the relative 
distance $d_r$, with $c_m$ and $d_r$ defined in (\ref{uno}) and  
(\ref{due}), where $x_1$ and $x_2$  
are the projections on the backbone of the positions of particles 1 and 2 
respectively.  
When one of the particles is on a finger, $c_m$ and $d_r$ have non integer 
values and $d_r$  has negative values when   
$x_1 > x_2$. The diffusion graphs of $c_m$ and $d_r$ are themselves narrow comb lattices, 
decorated with loops and triangles as in Fig. 3 and Fig. 4.   
The position of $c_m$ on the NCL can be easily deduced from the position of a random walker  
on the corresponding diffusion graph: when the random walker occupies a site on the  
backbone, its position corresponds to the value of $c_m$;  
when it stays on a finger, the value of $c_m$ is that of the position of 
that finger  
and both particles occupy sites on the fingers of the NCL. Finally,  
when the walker is on the upper vertex of one of the triangles,  $c_m$ has  
a semi-integer value corresponding to the projection of the vertex on the backbone.  
In this situation one of the particles is on the backbone and the other on the fingers.  
Loops are introduced on the diffusion graph  of the NCL to represent waiting time 
probabilities for $c_m$.\\  
The relative distance graph can be built in a similar way, taking   
contact into account. Point $O$ represents the tower configuration which can  
evolve on point A (contact with $x_1<x_2$) or in point A'  
(contact with $x_1>x_2$) and all the other points will be indicated with the  
value of their projection on the backbone ($1/2$, $1$, $3/2$, $2$...  
and $-1/2$, $-1$, $-3/2$, $-2$..., depending on the walker being on the left or the 
right side of the graph with respect to point $O$).\\  
The diffusion graph with the corresponding jumping probabilities is represented   
in Fig. 4. The jumping probabilities not explicitly indicated are equal to the  
corresponding ones of the $c_m$ diffusion graph.  
Since points A and A' represent contacts between particles, we can have 
particle crossing if  
A $\rightarrow $ A' or   A' $\rightarrow $ A i.e. crossing can only be the consequence  
of a contact.  
Now we can study the motion of the two interacting particles analyzing random walks on the  
diffusion graphs.\\  
The first quantity we calculate is the probability of a contact between particles  
without crossing. This corresponds to the probability that a single  
walker starting from point $A$ of the diffusion graph of $d_r$ returns to point $A$  
after $t$ steps without ever reaching point $A'$. We call this quantity  
$P^{(A')}(A,A;t)$ and we define $F^{(A')}(A,A;t)$ as the probability that a 
walker  starting from point $A$ returns to this same point for the first time 
after $t$ steps without ever reaching point $A'$. We have:  
\[  
	F^{(A')}(A,A;t)=\frac{1}{3} \frac{1}{2} \delta_{t,2}  
	+\frac{1}{6} \delta_{t,2} + +\frac{1}{9}\left[\frac{1}{6}P^{(A)}(1,1;t-2) +\right.  
\]  
\begin{equation}  
	 \left. +  
	\frac{1}{3} P^{(A)} \left(\frac{1}{2},1;t-2 \right) +  
	\frac{1}{2} P^{(A)}\left(1,\frac{1}{2},t-2\right)+  
	 P^{(A)} \left(\frac{1}{2},\frac{1}{2};t-2\right) \right]  
\end{equation}  
  
where the first term refers to the two steps $A$ $\rightarrow$ $O$  $\rightarrow$  
$A$, the second to the probability of reaching the finger corresponding to point $A$ and  
then going back to the backbone. $P^{(n)}(i,j;t)$ is the probability of a $t$ steps  
walk from point $i$ to point $j$ never reaching point $n$.  
Let us introduce the generating function of a generic $P^{(n)}(i,j;t)$ as:  
\begin{equation}  
\widetilde{P}^{(n)}(i,j;\lambda)=\sum_{t=0}^{\infty}P^{(n)}(i,j;t)  
\lambda ^t  
\end{equation}  
Using the relation $\widetilde{P}^{(n)}(i,i;\lambda)=  
1/[1-\widetilde{F}^{(n)}(i,i;\lambda)] $\cite{MeW}, the calculation of one of the  
$\widetilde{P}^{(n)}(i,i;\lambda)$ is reduced to that of the corresponding   
$\widetilde{F}^{(n)}(i,i;\lambda)$ which is, in general, a much easier   
task. In the case of $F^{(A')}(A,A;t)$ we obtain:  
\[  
		\widetilde{F}^{(A')}(A,A;\lambda)=  
		\frac{\lambda^2}{6} +  \frac{\lambda^2}{6}+  
\frac{\lambda^2}{9} \left[\frac{1}{6}  
	\widetilde{P}^{(A)}(1,1;\lambda) +\right.  
	\]  
\begin{equation}  
	\left. +\frac{1}{3 } \widetilde{P}^{(A)}\left(\frac{1}{2},1;\lambda \right) +  
		\frac{1}{2} \widetilde{P}^{(A)}\left({1,\frac{1}{2}};\lambda \right) +  
		 \widetilde{P}^{(A)} \left(\frac{1}{2},\frac{1}{2};\lambda \right)\right]  
	\end{equation}  
The generating functions of $\widetilde{P}^{(A)}(1,1;\lambda)$ and   
$\widetilde{P}^{(A)}(1/2,1/2;\lambda)$ can in turn be expressed as functions of the  
corresponding first time generating functions:  
\begin{equation}  
\widetilde{F}^{(A)}({1,1};\lambda) =  
\frac{2}{27}\lambda^2\frac{1}{1-\frac{\lambda}{3}}  
	+x  
	\label{}  
\end{equation}  
\begin{equation}  
\widetilde{F}^{(A)}\left({1 \over 2},{1 \over 2};\lambda \right) =  
{1 \over 3}\lambda + \frac{2}{27}\lambda^2\frac{1}{1-x}   
\end{equation}  
  
where:  
\[ x=\widetilde{F}^{(A, 1/2)}({1,1};\lambda)=  
\frac{\lambda^2}{9}+\frac{2}{9} \lambda+  
                \frac{\lambda^2}{9} \left[\frac{1}{9}  
                \widetilde{P}^{(1)}({2,2};\lambda)+ \right. \]  
\begin{equation}  
\left.                +\frac{2}{9}  
                 \widetilde{P}^{(1)}\left({\frac{3}{2},2};\lambda \right)  
                +\frac{1}{3}\widetilde{P}^{(1)}\left({2,\frac{3}{2}};\lambda\right)  
                +\frac{2}{3}\widetilde{P}^{(1)}  
                \left({\frac{3}{2},\frac{3}{2}};\lambda \right) \right]  
                \label{otto}  
\end{equation}  
Moreover since  
$\widetilde{P}^{(n)}(i,j;\lambda)=\widetilde{F}^{(n)}(i,j;\lambda) 
\widetilde{P}^{(n)}(j,j;\lambda) $ we also have:  
  
\begin{equation}  
 \widetilde{P}^{(A)}\left({\frac{1}{2},1};\lambda \right)  
={1 \over {1-{\lambda \over 3}}}\cdot {\lambda \over 3}\widetilde{P}^{(A)}\left(  
{1,1};\lambda \right)  
\end{equation}  
  
\begin{equation}  
 \widetilde{P}^{(A)}\left({1,\frac{1}{2}};\lambda \right)  
={1 \over {1- x}}\cdot {{2\lambda} \over {9}}\widetilde{P}^{(A)}\left(  
{1\over 2},{1\over 2};\lambda \right)  
\end{equation}  
Using the translation invariance in the backbone direction of the comb, 
we get the equalities:  
$\widetilde{P}^{(1)}(2,2;\lambda)=\widetilde{P}^{(A)}(1,1;\lambda)$;  
$\widetilde{P}^{(1)}(3/2,2;\lambda)=\widetilde{P}^{(A)}(1/2,1;\lambda)$;  
$\widetilde{P}^{(1)}(2,3/2;\lambda)=\widetilde{P}^{(A)}(1,1/2;\lambda)$ and  
$\widetilde{P}^{(1)}(3/2,3/2;\lambda)=\widetilde{P}^{(A)}(1/2,1/2;\lambda)$.  
We obtain from (\ref{otto}) an equation for $x$:  
\begin{equation}  
	 x = \frac{\lambda^2}{9}+\frac{2}{9} \lambda+ \frac{\lambda^2}{9}  
	 \left[ \frac{3+3\lambda+18(1-x)}{9(3-\lambda)(1-x)-2\lambda^2}  
	 \right]  
\end{equation}  
Now we can substitute the value of $x$ and obtain the final expression:  
\begin{equation}  
\widetilde{F}^{(A')}(A,A;\lambda)=  
\frac{\lambda^2}{6}+\frac{\lambda^2}{6}  
\left(1+\frac{3+3\lambda+18(1-x)}{9(3-\lambda)(1-x)-2\lambda^2}\right)  
\end{equation}  
The asymptotic behavior of $P^{(A')}(A,A;t)$ can be derived from that of the  
corresponding generating function considering the limit $\lambda \rightarrow 1$  
and then applying Tauberian theorems \cite{MeW}:
\begin{equation}  
     {P^{(A')}}(A,A;t) \sim 12 \sqrt{\frac{2}{\pi}}\mbox{}t^{-\frac{3}{2}}  
     \qquad t \to \infty
\label{u1}  
\end{equation}
This asymptotic law shows a dramatic difference with respect to the case 
of the simple linear chain where particle crossing is forbidden and the 
probability of a contact decays as $t^{-1/2}$. Notice that the deep 
change in the asymptotic behavior,
originated by the new geometry, has no analogous if we consider the 
motion of a single particle, since in this case the linear chain and the 
NCL have the same asymptotic diffusion laws.\\ 
\noindent In a similar way, we can calculate the probability of particles 
crossing.  
This is the probability $P(A,A';t)$ that a random walker starting from point $A$ of the  
diffusion graph of $d_r$ reaches point $A'$ after a $t$ steps walk. Since   
\begin{equation}  
       \widetilde{P}(A,A';\lambda)=\widetilde{F}(A,A';\lambda)   
       \widetilde{P}(A',A';\lambda)  
\end{equation}  
we find after some calculations:  
 \begin{equation}  
 	P(A,A';t) \sim  \frac{3}{8}\sqrt{\frac{2}{\pi}}\mbox{} t^{-\frac{1}{2}}  
         \qquad t \to  \infty 	  
\label{u2} 
\end{equation}  
A key quantity is the average number of contacts without particles crossing.  
This is given by the average number of returns to point $A$ after $t$ steps on the $d_r$  
diffusion graph without reaching point $A'$:  
\begin{equation}  
       	M^{(A')}(A,A;t)\sim const  
\end{equation}  
Finally, the mean number of particles crossing follows the asymptotic law  
\begin{equation}  
              M(A,A',t) \sim  \frac{3}{4}\sqrt{\frac{2}{\pi}}\mbox {}   
t^{\frac{1}{2}}   
              \qquad t \to \infty  
\end{equation}  
For a better description  of the diffusion properties of the two particles
on the NCL, we study the mean square dispersions $\Delta x_1 ^2$
and  $\Delta x_2 ^2$, defined as
\begin{equation} 
\Delta x_1 ^2= \Delta x_2 ^2= <x_1^2>-<x_1 >^2=<x_2^2>-<x_2 >^2
\end{equation}
In \cite{Asl1} the  mean square dispersions of two hard core 
interacting particles on a one dimensional lattice was studied. 
In this case, particle crossing was 
forbidden by the  potential and by the geometry of the system and it 
was found that 
 \begin{equation} 
\Delta x_1 ^2= \Delta x_2 ^2= \left( 1 -{ 1\over \pi}\right)c\,t
\end{equation}
$c$ being the diffusion constant, while for a single particle moving freely on a one dimensional lattice
we have
 \begin{equation} 
\Delta x_1 ^2=c\, t
\end{equation}
From the previous results it follows that the interaction  inhibits through
the factor $(1- 1/\pi)$ the spreading. In the NCL case, from (\ref{uno}) 
and 
 (\ref{due}) it follows that:
 \begin{equation} 
<x_1>=<c_m-d_r>   
\label{5.1}
\end{equation}
\begin{equation} 
<x_2>=<c_m+d_r>+1
\label{5.2}
\end{equation}
The expressions of $<c_m>$ and $<d_r>$, due to the 
{\it{diffusion graph transform}}, can be calculated as the mean 
displacement $<d>$ of a single walker moving on the correspondent 
diffusion graph, $d$ being the coordinate of the projection of the position  
of the walker on the backbone of the diffusion graphs.
Using the equation 
\begin{equation} 
<d>={{\sum_{-\infty}^{\infty}d\, P(O,d;t)
}\over {\sum_{-\infty}^{\infty} P(O,d;t)}}
\end{equation} 
and following the steps described in \cite{Bethe} we find 
$<c_m>=<d_r>=0$: this also immediately follows 
from the fact that the diffusion graphs of $c_m$ and $d_r$ are symmetric 
with respect to point $O$ so that $<c_m>$ and $<d_r>$ must be $0$.
Equations (\ref{5.1}) and (\ref{5.2}) become 
\begin{equation}
<x_1>=<c_m-d_r>=0   
\label{5.3}
\end{equation}
\begin{equation} 
<x_2>=<c_m+d_r>+1=1
\label{5.4}
\end{equation}
and we have also 
\begin{equation} 
\Delta x_1 ^2=\Delta x_2 ^2=<c_m^2>+<d_r^2>
\label{5.5}
\end{equation}
By straightforward calculations we find 
\begin{equation} 
<c_m^2>\sim {1\over 4} t \qquad  t \to \infty 
\end{equation} 
\begin{equation} 
<d_r^2>\sim {1\over 4} t \qquad t \to \infty 
\end{equation}
so that 
\begin{equation} 
\Delta x_1 ^2=\Delta x_2 ^2 \sim {1\over 2} t \qquad t \to \infty   
\label{5.6}
\end{equation}
For a single particle diffusing on the NCL we have
\begin{equation} 
\Delta x ^2\sim {1\over 2} t \qquad t \to \infty   
\end{equation}
where $x$ is defined as the projection of the particle position on the
 backbone of the lattice  (for a single walker on a linear chain 
$\Delta x ^2\sim  t \quad  t \to \infty$).
We can therefore conclude that the interaction between particles does not
modify the spreading as a consequence of the possibility of particle crossing.
This result is confirmed by equations (\ref{u1}) and (\ref{u2}),
stating that a contact with crossing is much more frequent than a 
contact without crossing.

\section{Two interacting particles on ladder lattices }  
The ladder lattice exhibits new features with respect to the previous cases  
due to the presence of loops. The ladder graph is made of two linear  
chains whose corresponding points are connected by links as in Fig. 2 b.  
The two particles start from the same site and then move on the ladder with  
jumping probabilities equal to $1/3$. The relative distance is defined as $d_r=N/2$,  
$N$ being the chemical distance i.e. the number of links of the shortest   
path connecting the two particles. The $d_r$ diffusion graph is represented in Fig. 5;  
the value of $d_r$ is given by the projection on the lower chain   
of the position of a random walker.  When the walker is on the lower chain  
of the $d_r$ diffusion graph, particles on the real ladder lattice are on the 
same chain,  
while when the walker jumps on the upper chain of the $d_r$ diffusion   
graph it means that particles move on different chains. Point $O$   
represents the contact between particles and the link   
between $A$ and $A'$ represents the possibility of particles crossing each   
other without direct contact, since it connects points with $d_r>0$ and 
$d_r<0$ 
without passing  through point $O$. This is the direct consequence
 of the presence 
of loops and it represents a fundamental technical difference with respect to 
the previous case.\\
The solution of the random walks problem on the diffusion graph can be
simplified by the following considerations. As already shown in \cite{India} 
the diffusion problem on a ladder-like lattice can be mapped into
that of a linear chain. This follows from the fact that we are interested only 
in the walker  displacement in the unbounded direction, and not in its 
position on 
the upper or on the lower chain. Then the ladder can be considered as  
a linear chain where we define  additional  staying probabilities 
corresponding to the steps spent for jumping from one chain to the other. 
In the present case this simplification can be applied to the two half-linear 
chains originating from the central loop. Notice that the central loop 
cannot be reduced to a couple of points of the linear chain and must survive 
our transformation since it breaks the translational symmetry of the network.
This detail is crucial for the calculation of crossing without contact 
probability.\\
The final graph obtained by this analysis is represented in Fig. 6. 
It reproduces the main features of the  
$d_r$ diffusion graph for the probability of particles crossing with or
without contact. From calculations analogous to the previous case we find that
the probability of contact without crossing is:  
\begin{equation}  
 P^{(A')}(A,O;t) \sim  
{36\over 25}\sqrt{ {2 \over \pi}}  t^{-\frac{3}{2}}  \qquad      t \to \infty  
\label{a}  
\end{equation}  
and shows the same time dependence we found in the NCL case. 
The probability of crossing without contact is:  
\begin{equation}  
P^{(O)}(A,A';t) \sim \frac{48}{25} \sqrt{{2 \over \pi}}
t^{-\frac{3}{2}}  \qquad        
t \to \infty  
\label{c}  
\end{equation}     
while the generic probability of crossing decays as:  
\begin{equation}  
 		P(A,A';t) \sim  
{3 \over {\sqrt{2 \pi}}} t^{-\frac{1}{2}}  \qquad      
t \to \infty  
 		\label{b}  
\end{equation}  
\noindent
The description of the steps leading to (\ref{a}), (\ref{c}), (\ref{b})
can be found in Appendix A.\\
We  see that the probability of a contact without crossing on the ladder
lattice follows the same law we found in the NCL case and this is  also true 
for the crossing probability.  On the ladder lattice however, a 
probability of crossing without contact can be defined. 
This particular feature of the ladder lattice can be extremely useful 
when the two processes representing {\it{contact}} and   
{\it{crossing}} are used to represent different physical situations.

\section{Summary and Discussion}  
The diffusing graph transform introduced in this paper is a powerful 
technique which, combined with the universality properties of the
long time asymptotic behavior of random walks return probabilities,
provide a new tool for the analysis of interacting diffusing particles
on discrete structures.  
The technique can be applied to decouple and analytically solve
the problem on tree-like and quasi $1$-dimensional 
structures also in presence of loops and crossing probabilities 
for the particles. This has been done for
the case of two interacting particles 
on the Narrow Comb and on the Ladder lattice. Our solution shows that 
the asymptotic behavior of contact probabilities is sensitive 
to the local geometrical details of the underlying discrete structure.

\section*{Appendix A }
\setcounter{equation}{0}
\renewcommand{\theequation}{A.\arabic{equation}}
The explicit calculation of  (\ref{a}), (\ref{c}), (\ref{b}) is based on two
properties of the generating functions \cite{MeW}. The first states that the generating 
function of the probability $\widetilde{P}^{(k)}(i,j;\lambda)$ of going 
from point $i$ to point $j$ ($i \neq j$), without ever touching point $k$, 
is related to the corresponding first time arrival generating function by the 
relation
\begin{equation}
 \widetilde{P}^{(k)}(i,j;\lambda)=\widetilde{F}^{(k)}(i,j;\lambda)\cdot
\widetilde{P}^{(k)}(j,j;\lambda)
\label{appendice1}  
\end{equation}
The second property is the well known relation between 
$\widetilde{P}^{(k)}(i,i;\lambda)$ and $\widetilde{F}^{(k)}(i,i;\lambda)$
we already used in the preceding sections;
\begin{equation}
 \widetilde{P}^{(k)}(i,i;\lambda)={ 1 \over {1 -\widetilde{F}^{(k)}(i,i;\lambda)}}
\label{appendice2}  
\end{equation}  
Let us start from the calculation of $\widetilde{P}^{(A')}(A,0;\lambda)$:
from (\ref{appendice1}) it follows that:
\begin{equation}
 \widetilde{P}^{(A')}(A,0;\lambda)=\widetilde{F}^{(A')}(A,0;\lambda)\cdot
\widetilde{P}^{(A')}(0,0;\lambda)
\end{equation}
since:
\begin{equation}
\widetilde{F}^{(A')}(A,0;\lambda)={\lambda \over 6}\widetilde{P}^{(0,A')}
(A,A;\lambda)
\end{equation}
and
\begin{equation}
\widetilde{P}^{(A')}(0,0;\lambda)={1 \over{1-\widetilde{F}^{(A')}(0,0;\lambda)}} = {1 \over{1-{\lambda^2 \over 12}\widetilde{P}^{(O,A')}(A,A;\lambda)}} 
 \end{equation} 
so that the calculation of $\widetilde{P}^{(A')}(A,0;\lambda)$ is reduced to that of
$\widetilde{P}^{(O,A')}(A,A;\lambda)$.
To obtain the expression of $\widetilde{P}^{(O,A')}(A,A;\lambda)$
we calculate the corresponding first time arrival generating function:
\begin{equation}
\widetilde{F}^{(O,A')}(A,A;\lambda) = { {5 \lambda} \over 9}+\left(
 { {2\lambda}\over 9} \right)^2 \widetilde{P}^{(A)}(1,1;\lambda)
 \end{equation} 
From the translational invariance of the lattice, we have that 
$\widetilde{P}^{(A)}(1,1;\lambda)=\widetilde{P}^{(O,A')}(A,A;\lambda)$.
Using the preceding relation together with (\ref{appendice2})
we obtain:
\begin{equation}
\widetilde{P}^{(O,A')}(A,A;\lambda){{ {1- {{5\lambda }\over 9} }-\sqrt{1+
{{\lambda^2 }\over 9 }-{{10\lambda }\over 9 }}}
\over { {8\lambda ^2}\over 81}}
\end{equation} 
In the limit $\lambda  \rightarrow 1- \epsilon$ and $\epsilon \rightarrow 0$
we have:
\begin{equation}
\widetilde{P}^{(O,A')}(A,A;1-\epsilon) \sim {9 \over 2}\left( 1- 3\sqrt{\epsilon  \over 2} \right)
\end{equation} 
and:
\begin{equation}
\widetilde{P}^{(  A')}(A,0;1-\epsilon) \sim {6 \over 5} - {72 \over 25} \sqrt{2
\epsilon}
\label{appendice3}
\end{equation}
Applying Tauberian theorems we obtain (\ref{a}).
For the determination of $\widetilde{P}^{(O)}(A,A';\lambda)$, 
we follow the same technique writing this quantity as a function of 
$\widetilde{P}^{(O,A')}(A,A;\lambda)$, taking the limit $\lambda  
\rightarrow 1- \epsilon (\epsilon \to0)$ and  using  Tauberian theorems.
In conclusion we have:
\begin{equation}
 \widetilde{F}^{(O)}(A,A';\lambda)={\lambda \over 18}\widetilde{P}^{(O,A')}(A,A;\lambda)
\end{equation}
\begin{equation}
 \widetilde{F}^{(O)}(A',A';\lambda)=
1-{ 1\over {\widetilde{P}^{(O,A')}(A,A;\lambda)}}+\left(  
{ \lambda\over 18}\right)^2 \widetilde{P}^{(O,A')}(A,A;\lambda)
\end{equation}
and finally using (\ref{appendice2}) we have 
 \begin{equation}
 \widetilde{P}^{(O)}(A,A';\lambda)=\widetilde{F}^{(O)}(A,A';\lambda)\cdot
\widetilde{P}^{(O)}(A',A';\lambda)
\end{equation}
For $\widetilde{P}(A,A';\lambda)$ the corresponding steps can be resumed as:
\begin{equation}
\widetilde{P}(A,A';\lambda)=\widetilde{F}(A,A';\lambda)
\widetilde{P}(A',A';\lambda)
\end{equation}
\[
\widetilde{F}(A',A';\lambda)=
1 - { 1\over {\widetilde{P}^{(O,A')}(A,A;\lambda)  }}+
\]
\begin{equation}
\frac{ \lambda ^2}{12 - \lambda ^2 \widetilde{P}^{(O,A')}(A,A;\lambda)}
\left( { 1\over 27}\widetilde{P}^{(O,A')}(A,A;\lambda)+
{ \lambda \over 9}\widetilde{P}^{(O,A')}(A,A;\lambda)+1
\right)
\end{equation}
\begin{equation}
\widetilde{F}(A,A';\lambda)=
\frac{\widetilde{P}^{(O,A')}(A,A ;\lambda)}  {
 12 - \lambda ^2     \widetilde{P}^{(O,A' )}(A,A ;\lambda)  } 
\left( {{ 2 \lambda      }\over 3 } + \lambda ^2 \right)
\end{equation}

\newpage

\centerline{Figure captions}  
\vspace{1cm}  
\noindent {\bf Fig.1}\\  
The diffusion graphs of a linear chain: a) $c_m$ diffusion graph and  
b) $d_r$ diffusion graph \\  
{\bf Fig.2} \\  
a) `Truncated comb lattice' b) `Ladder lattice'\\  
{\bf Fig.3} \\  
 $c_m$ diffusion graph and jumping probabilities for   
 the truncated comb lattice\\  
{\bf Fig.4} \\  
$d_r$ diffusion graph and jumping probabilities for   
 the `Truncated comb lattice'\\  
{\bf Fig.5} \\  
$d_r$ diffusion graph and jumping probabilities for   
 the `Ladder lattice'\\  
{\bf Fig.6} \\  
Semplified $d_r$ diffusion graph for the Ladder lattice\\

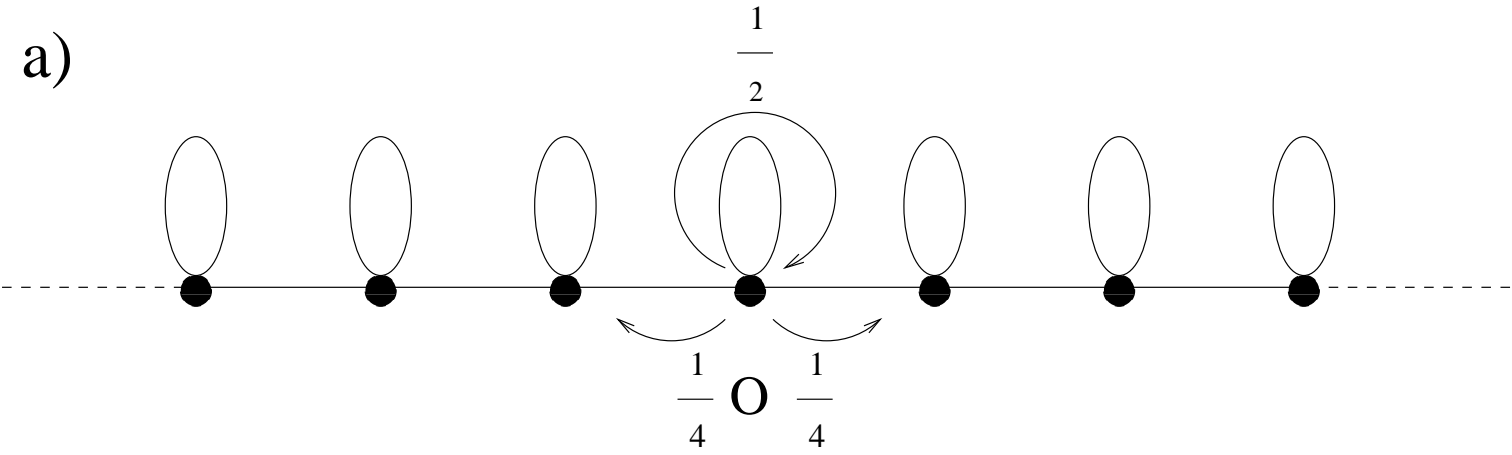
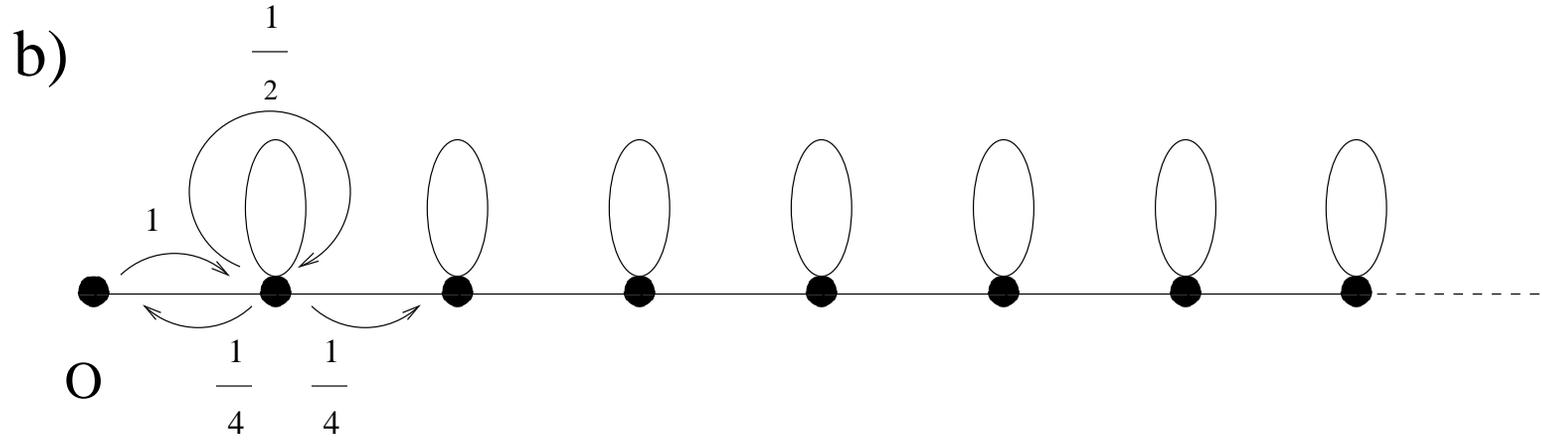
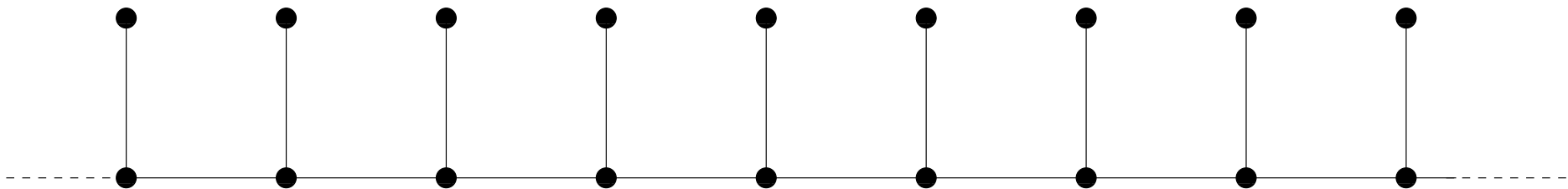
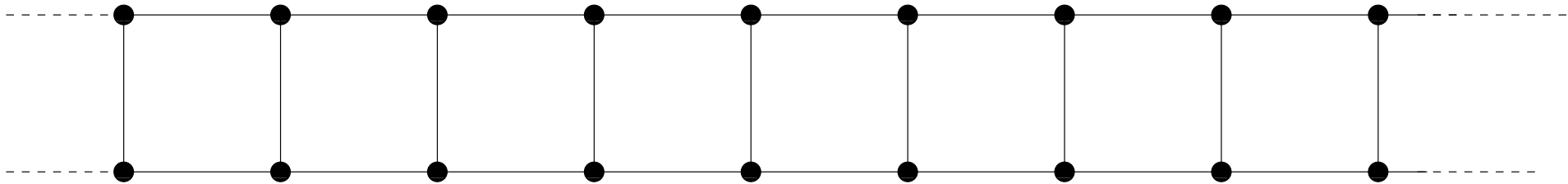
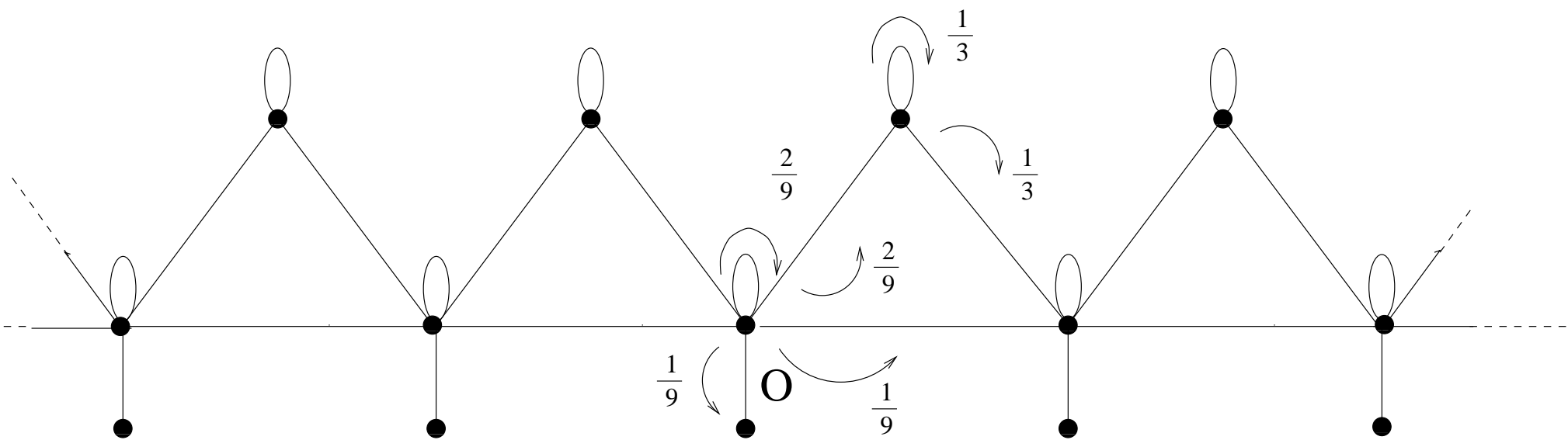
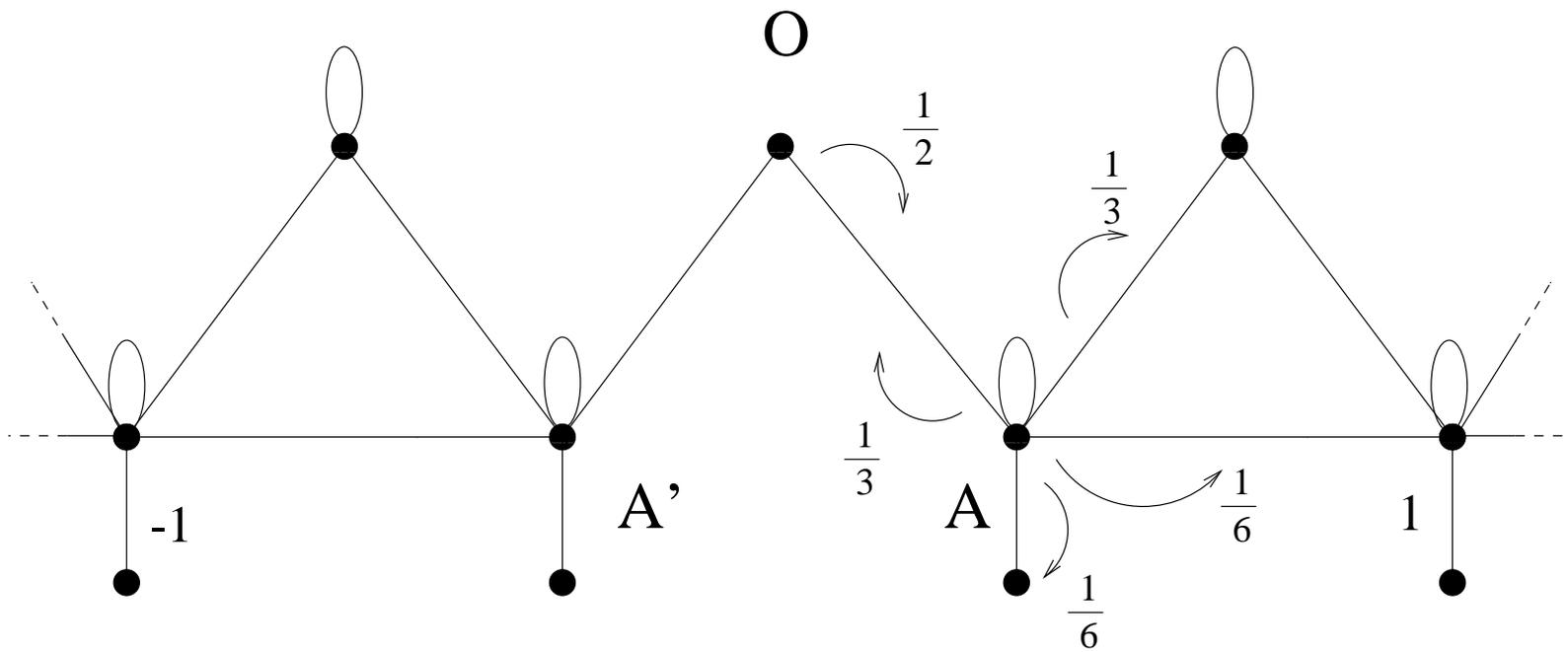
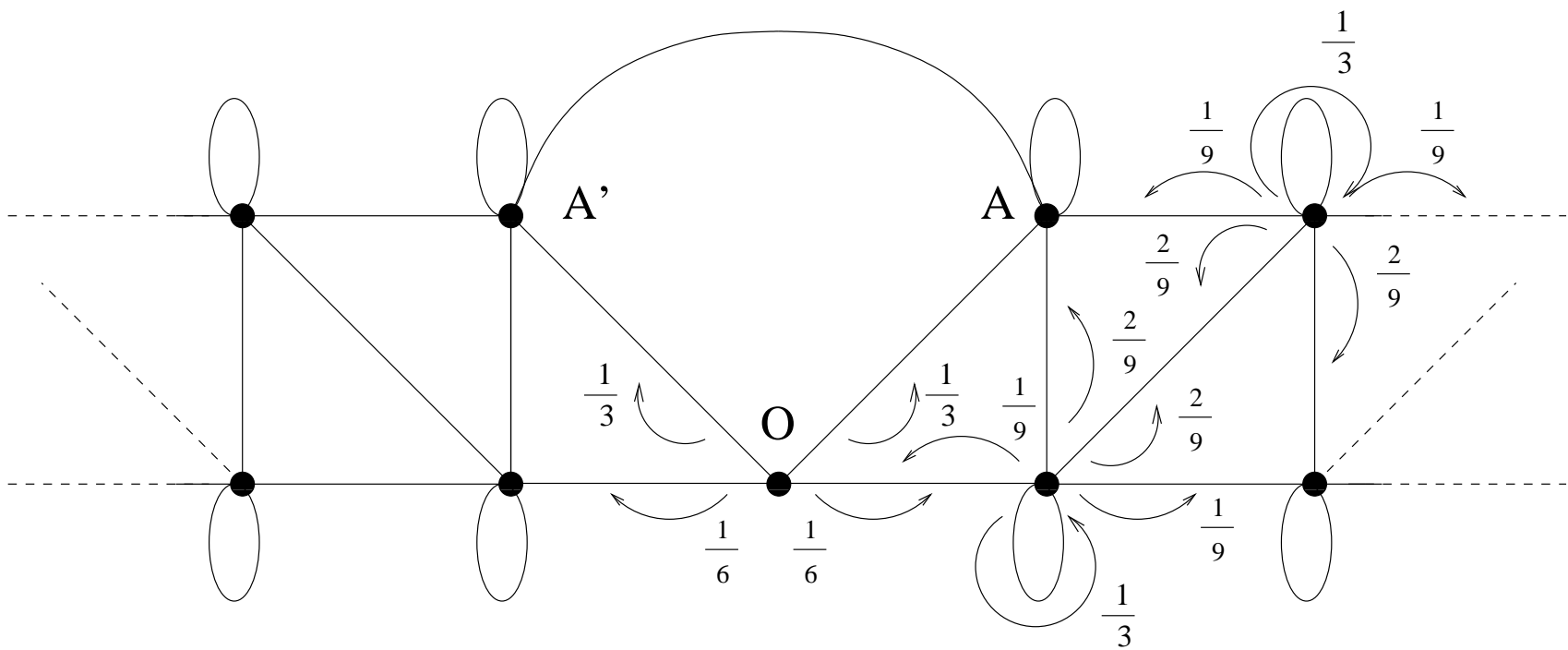
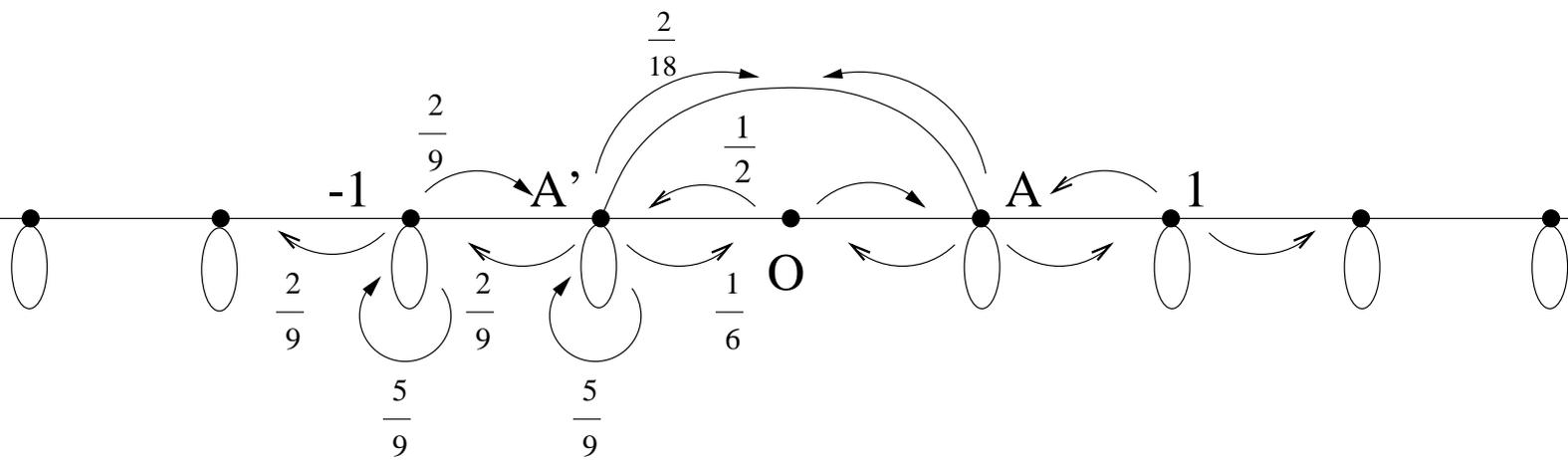
\end{document}